\documentclass[modern]{aastex61}

\usepackage{graphicx}

\usepackage{lineno}

\newcommand{\asec}{\hbox to 1pt{}\rlap{$^{\prime\prime}$}.\hbox to 2pt{}}
\newcommand{\amin}{\hbox to 1pt{}\rlap{$^{\prime}$}.\hbox to 1pt{}}
\newcommand{\adeg}{\hbox to 1pt{}\rlap{$^{\circ}$}.\hbox to 2pt{}}

\shortauthors{Lauer et al.}
\shorttitle{The Dark Side of Pluto}

\begin{document}

\title{The Dark Side of Pluto}

\author{Tod R. Lauer}
\affil{NSF's National Optical Infrared Astronomy Research
Laboratory,\footnote{The NSF's OIR Lab is operated by AURA, Inc under cooperative agreement with NSF.} P.O. Box 26732, Tucson, AZ 85726}

\author{John R. Spencer}
\affil{Department of Space Studies, Southwest Research Institute, 1050 Walnut St., Suite 300, Boulder, CO 80302}

\author{Tanguy Bertrand}
\affil{LESIA, Observatoire de Paris, Universit\'e PSL, CNRS, Sorbonne Universit\'e, Universit\'e de Paris, 5 place Jules Janssen, 92195 Meudon, France}
\affil{NASA Ames Research Center, Space Science Division, Moffett Field, CA 94035}

\author{Ross A. Beyer}
\affil{Sagan Center at the SETI Institute, NASA Ames Research Center, United States, 95043}
\affil{NASA Ames Research Center, Space Science Division, Moffett Field, CA 94035}

\author{Kirby D, Runyon}
\affil{The Johns Hopkins University Applied Physics Laboratory, Laurel, MD 20723-6099}

\author{Oliver L, White}
\affil{Sagan Center at the SETI Institute, NASA Ames Research Center, United States, 95043}
\affil{NASA Ames Research Center, Space Science Division, Moffett Field, CA 94035}

\author{Leslie A. Young}
\affil{Department of Space Studies, Southwest Research Institute, 1050 Walnut St., Suite 300, Boulder, CO 80302}

\author{Kimberly Ennico}
\affil{NASA Ames Research Center, Moffett Field, CA 94035}

\author{William B. McKinnon}
\affil{Department of Earth and Planetary Sciences and McDonnell Center for the Space Sciences, Washington University, St. Louis, MO 63130}

\author{Jeffrey M. Moore}
\affil{NASA Ames Research Center, Space Science Division, Moffett Field, CA 94035}

\author{Catherine B. Olkin}
\affil{Department of Space Studies, Southwest Research Institute, 1050 Walnut St., Suite 300, Boulder, CO 80302}

\author{S. Alan Stern}
\affil{Space Science and Engineering Division, Southwest Research Institute, 1050 Walnut St., Suite 300, Boulder, CO 80302}

\author{Harold A. Weaver}
\affil{The Johns Hopkins University Applied Physics Laboratory, Laurel, MD 20723-6099}

\begin{abstract}

During its departure from Pluto, New Horizons used its LORRI camera to image a portion of Pluto's southern hemisphere that was in a decades-long seasonal winter darkness, but still very faintly illuminated by sunlight reflected by Charon. Recovery of this faint signal was technically challenging. The bright ring of sunlight forward-scattered by haze in the Plutonian atmosphere encircling the nightside hemisphere was severely overexposed, defeating the standard smeared-charge removal required for LORRI images. Reconstruction of the overexposed portions of the raw images, however, allowed adequate corrections to be accomplished. The small solar elongation of Pluto during the departure phase also generated a complex scattered-sunlight background in the images that was three orders of magnitude stronger than the estimated Charon-light flux (the Charon-light flux is similar to the flux of moonlight on Earth a few days before first quarter).  A model background image was constructed for each Pluto image based on principal component analysis { (PCA)} applied to an ensemble of scattered-sunlight images taken at identical Sun-spacecraft geometry to the Pluto images.  The recovered Charon-light image revealed a high-albedo region in the southern hemisphere. We argue that this may be a regional deposit of ${\rm N_2}$ or ${\rm CH_4}$ ice. The Charon-light image also shows that the south polar region currently has markedly lower albedo than the north polar region of Pluto, which may reflect the sublimation of ${\rm N_2}$ ice or the deposition of haze particulates during the recent southern summer.
\end{abstract}


\section{Using Charon to See in the Dark}

As NASA's New Horizons spacecraft departed Pluto following its July 14, 2015 flyby encounter \citep[summarized in][]{stern}, the planet presented almost its entire nightside hemisphere for observation. Departure imaging from this geometry provided sensitive measurements of the haze content of the Pluto atmosphere, due to its enhanced visibility in forward-scattered sunlight \citep{gladstone, cheng}.  Deep searches for faint dust clouds or rings around Pluto were also conducted during this time \citep{lauer}. Finally, an attempt was made to detect ``Charon-light" illumination of the nightside hemisphere of Pluto by sunlight reflected off of Pluto's moon Charon. At the time, Pluto was experiencing its northern late spring \citep{binzel}, immersing latitudes south of $-39^\circ$ into seasonal darkness.\footnote{Strong twilight illumination provided by Pluto's hazy atmosphere remains significant for a latitudinal zone south of this limit \citep{moore}.} Charon-light imaging thus offered the potential to reveal surface features not accessible by other means in a portion of the southern hemisphere.  Charon-light imaging also offered a probe of seasonal albedo variations, due to the sublimation or deposition of surface ice, by imaging zones that had recently transitioned {into seasonal darkness after a long interval of daily solar illumination.}

The Charon-light observations were attempted in the P\_DEEPIM imaging sequence, which used New Horizons' Long Range Reconnaissance Imager (LORRI; see \citealt{lorri}) to  obtain 360 deep images of Pluto's nightside hemisphere over a 19m interval (and followed by an equal number of scattered-sunlight calibration images).  With an asymptotic departure sub-spacecraft latitude of $-43^\circ,$ a large portion of the southern hemisphere that had Charon in the sky was potentially detectable after closest approach if the illumination from Charon was strong enough.  Charon orbits Pluto synchronously with an inclination of $0\adeg0$ with respect to Pluto's equator \citep{orbit}. The sub-Charon point on the equator in fact defines the prime meridian of Pluto's longitude system. The timing of the New Horizons encounter was phased for optimal viewing of Sputnik Planitia,\footnote{The feature now designated as Sputnik Planitia was evident as a large high-albedo feature in the \citet{buie} HST high-resolution maps prepared in advance of the encounter, and was also known to correspond to the Planet's light-curve maximum and a region of anomalously high CO abundance.} which is opposite Charon at longitude $180^\circ.$  As such, the majority of the hemisphere illuminated by Charon was in darkness at the time of the encounter.

\begin{figure}[thbp]
\centering
\includegraphics[keepaspectratio,width=6.0 in]{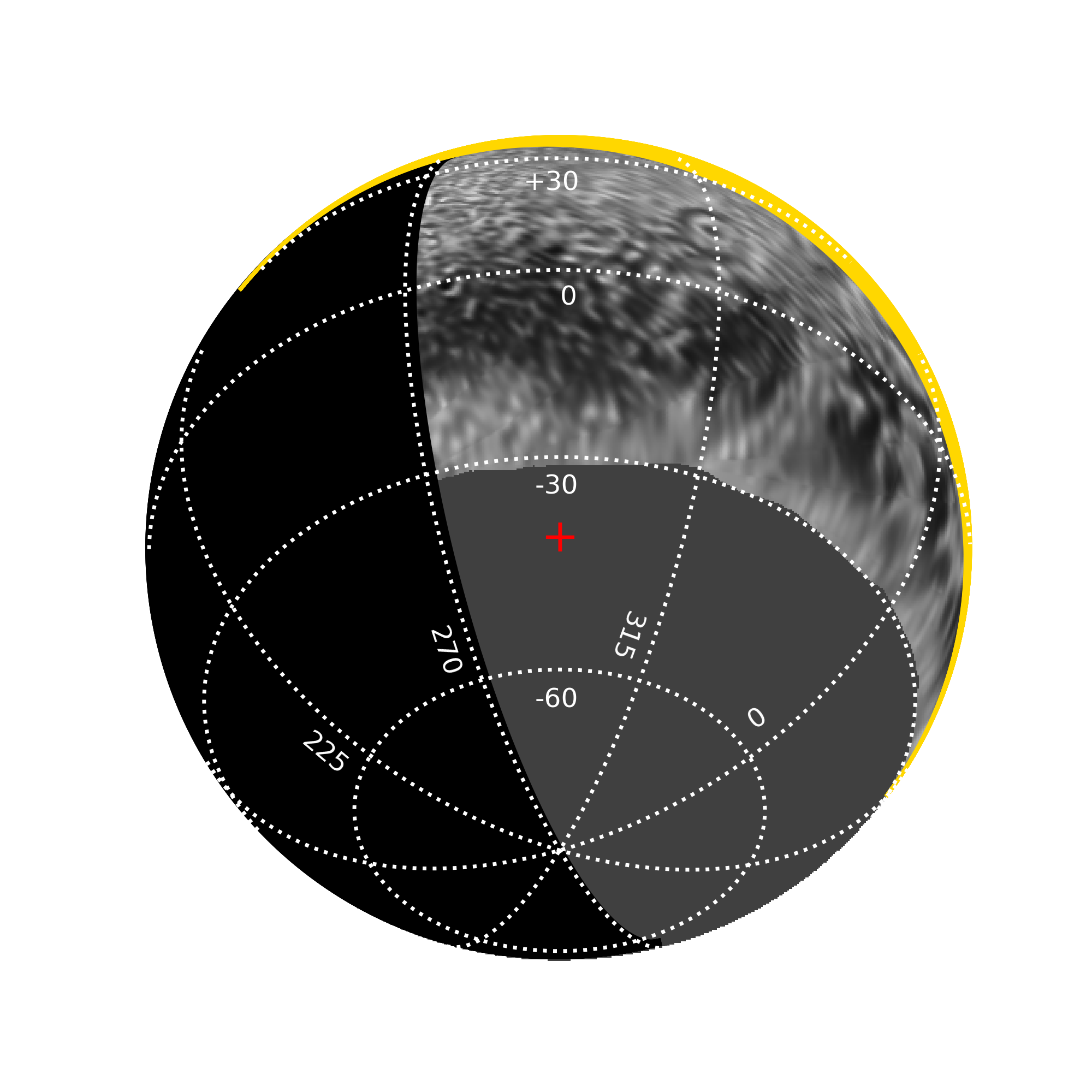}
\caption{The viewing geometry of Pluto during the P\_DEEPIM sequence is shown. The sub-spacecraft position on the planet is at the center of the disk {(marked with the red $+$)}. The south pole of Pluto is near the bottom of the disk.  Lines  of latitude are given in $30^\circ$ increments.  Longitude is given in $45^\circ$ increments.  The sub-Charon point is at the right edge of the disk where the $0^\circ$ meridian and equator intersect. Charon illuminates a portion of the planet that was on the ``farside" during the encounter, and was thus imaged  at low resolution \citep{farside}.  The dark-gray sector marks the area of Pluto in seasonal darkness but still illuminated by  Charon-light during the encounter. Portions of  Pluto illuminated by sunlight are yellow and are overexposed in the P\_DEEPIM images.  Zones with neither Charon nor sunlight illumination are black.}
\label{fig:pluto_geo}
\end{figure}

The geometry of the P\_DEEPIM imaging is shown in Figure \ref{fig:pluto_geo}. During the sequence, the sub-spacecraft position on Pluto was at longitude $292\adeg1$ and latitude $-43\adeg3.$ The phase angle on Pluto with respect to the Sun was $166\adeg1,$ which meant that the illuminated crescent extended only $13\adeg9;$ given the strong foreshortening, this was visible as only a small sliver of illumination along the limb. The angle between Charon and the spacecraft with Pluto at the vertex was $75\adeg9,$ thus Charon-light illuminated slightly more than half of the departure hemisphere visible to New Horizons. Since Charon is positioned above  Pluto's  equator, P\_DEEPIM could potentially recover information at nearly all southern latitudes over a range of longitudes\footnote{{Charon is fully below} the horizon on Pluto south of latitude $-88.0^\circ.$}.
\clearpage
\subsection{The Estimated Charon-Light Strength}

\begin{figure}[htbp]
\centering
\includegraphics[keepaspectratio,width=6.0 in]{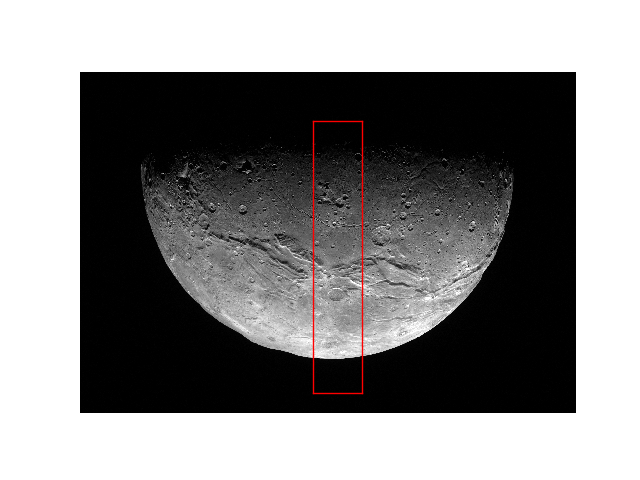}
\caption{The C\_MVIC\_LORRI\_CA  MVIC image of Charon {(MET 0299180333)} is shown, which is used to estimate the Charon-light flux on Pluto. The solar phase at this vantage point is $83\adeg7,$ which is only slightly less than the $88\adeg0$ solar phase of Charon during the P\_DEEPIM sequence. The red box marks the location of the high-resolution LORRI strip of images taken in parallel with the MVIC image, and which are used to calibrate it.}
\label{fig:c_mvic}
\end{figure}

We estimated the expected strength of Charon-light on Pluto from encounter images of Pluto and Charon selected to have solar-illumination phases similar to the solar-illumination phase of Charon and the Charon-illumination phase of Pluto in the P\_DEEPIM sequence.
Using observed image brightnesses directly, with modest corrections, rather than relying entirely on photometric model fits to the image data, greatly reduced the model dependence of our inferred Charon-light illumination.
Calculation of the flux from Charon strongly benefited from the the C\_MVIC\_LORRI\_CA  MVIC (the Multispectral Visible Imaging Camera on New Horizons) image of Charon, which was observed at nearly the same solar phase and Charon longitude at which Charon illuminated Pluto during the P\_DEEPIM sequence (Figure \ref{fig:c_mvic}). As this image was  being obtained, Charon was also imaged in parallel with the C\_MVIC\_LORRI\_CA  LORRI sequence, which obtained a strip of high-resolution images across the projected meridian of Charon. The  detected flux  in these images could  be  used to calibrate the integrated flux provided by the full disk of Charon. 

\begin{figure}[thbp]
\centering
\includegraphics[keepaspectratio,width=6.0 in]{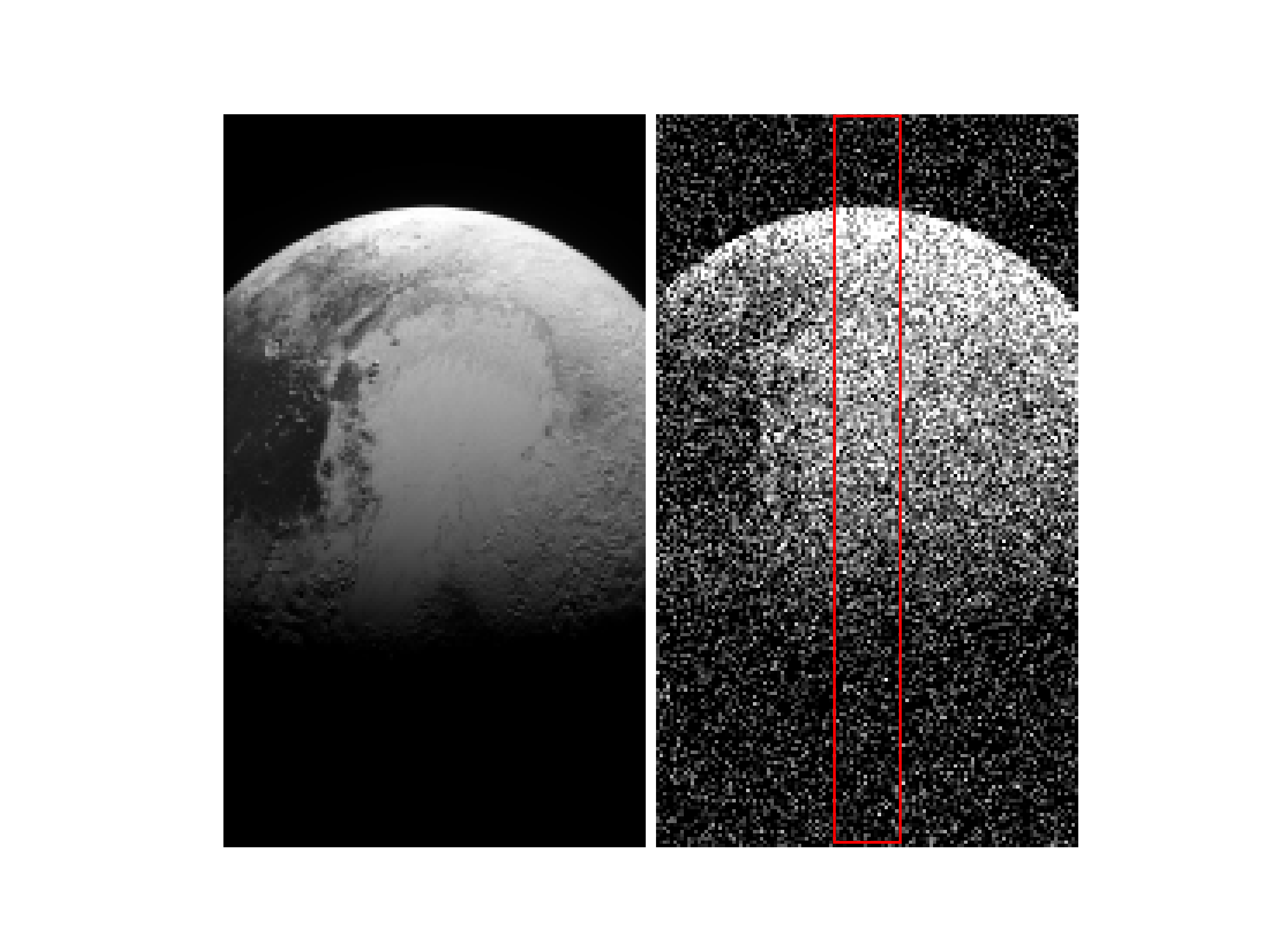}
\caption{The P\_MVIC\_LORRI\_CA  MVIC scan (MET 0299179552) of Pluto taken at close approach was observed at a solar phase-angle of $71\adeg3,$  making it an excellent proxy with appropriate scaling to simulate the P\_DEEPIM LORRI images.  The image on the left shows the MVIC image binned to match the P\_DEEPIM physical resolution.  The image to the right shows the simulated noise level for the final Charon-light image produced by stacking all 360 P\_DEEPIM images. The red box shows the area, 20 pixels in width, used to compute the trace in Figure \ref{fig:cut}.}
\label{fig:p_mvic}
\end{figure}

From the calibration provided by the LORRI images applied to the full MVIC Charon image, we estimated the $I/F$  ratio of solar flux returned from Charon to that incident on Charon to be 0.059, where this was an average over the full circular disk of Charon, including the half of the disk that was in darkness at this viewing geometry. The phase angle of the C\_MVIC\_LORRI\_CA is $83\adeg7,$ while the Charon phase as viewed from Pluto is $88\adeg0$ during the P\_DEEPIM sequence. From the Charon phase-curve measured by \citet{buratti}, the integrated flux decreases by 0.04 mag/degree at this part  of the curve, giving $I/F=0.050$ during the P\_DEEPIM sequence. The ratio, $R,$ of the Charon-light flux delivered to Pluto relative to the incident solar flux is then determined  by the solid angle, $\Omega,$ that Charon subtends as seen from Pluto, such that $R=(I\Omega)/(\pi F).$ With the 606 km radius of Charon \citep{nimmo}, and the Charon-Pluto distance of 18,997 km,\footnote{{This is the distance to a point halfway between the center of Pluto and the sub-Charon point on its surface} to represent the typical distance to Charon over the illuminated hemisphere.}  $\Omega =3.2\times10^{-2}$ sr, giving $R=5.1\times10^{-5}.$\footnote{In passing, we note that the calculated $R$ shows that the illumination on Pluto provided by Charon-light at any phase is about half of the moonlight flux on Earth provided by the Moon at the same phase. In short, the much larger solid angle of Charon as seen from Pluto, and its higher albedo compared to the Moon, nearly counter the strongly diminished solar flux at Pluto. For the present P\_DEEPIM sequence, the light provided by Charon is  about that provided by the Moon a day or so prior to first quarter.}

\begin{figure}[htbp]
\centering
\includegraphics[keepaspectratio,width=6.0 in]{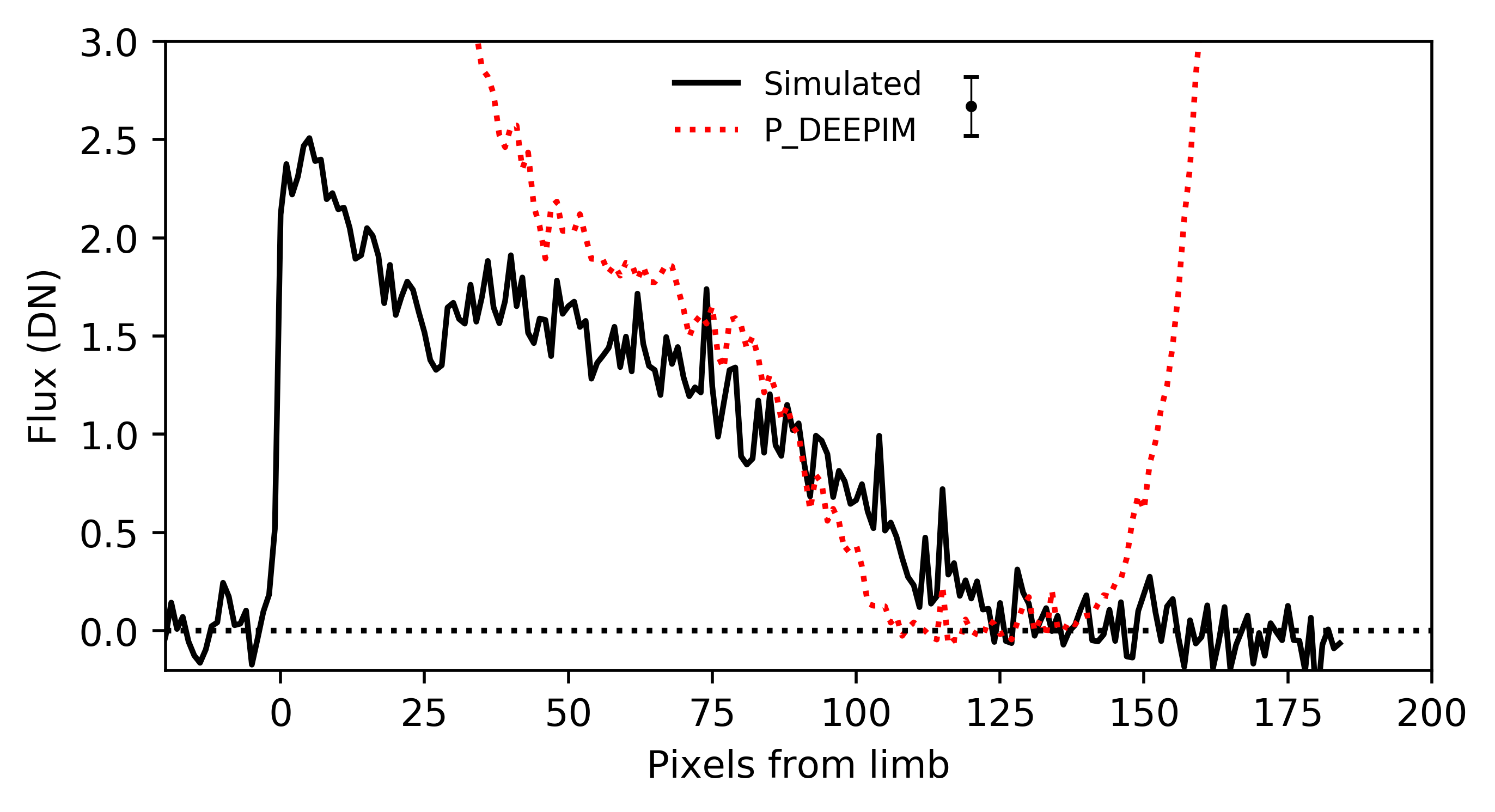}
\caption{The black trace shows the simulated flux as a function of distance from the limb, drawn from the cut shown in Figure \ref{fig:p_mvic}. The dotted red trace shows an actual intensity cut from the final P\_DEEPIM stack, and is discussed in $\S\ref{sec:ds_image}.$ The point to the right of the legend shows the average flux error for both traces, which are based on slices 20 pixels in width. {The real P\_DEEPIM trace rises at both ends owing to the bright haze ring surrounding Pluto, which was not included in the simulated trace.}}
\label{fig:cut}
\end{figure}

Having calculated the flux of Charon-light  relative to sunlight at Pluto, the next step was to estimate the observed brightness of Pluto in Charon-light as seen by LORRI. The simplest approach was to identify sunlight-illuminated images of Pluto obtained at solar phase-angles similar to that of the Charon illumination in the present Charon-light images, and scale them by the $R$ factor just derived.  The P\_MVIC\_LORRI\_CA MVIC-image (MET 0299179552) covered most of the disk of Pluto, was obtained at a phase angle of $71\adeg3,$ and thus was an excellent proxy for simulations of the Charon-light LORRI images, which have a Charon-illumination phase angle of $75\adeg9.$ With the estimated Charon-light and scaling the MVIC image to a $4\times4$ 0.3s LORRI exposure, the observed LORRI image DN {(data number value)} was scaled down by $1.30\times10^{-3}$ of the DN in the sunlight-illuminated MVIC image.  

The MVIC image of Pluto is presented in the left panel of Figure \ref{fig:p_mvic} binned to the same physical resolution as the P\_DEEPIM images. The peak brightness in this rendition was only 2.5 DN, which occurred at the limb. {In comparison, the typical scattered-sunlight background-level in the P\_DEEPIM images, was $\sim1600$ DN, which produces a shot-noise level of 9.0 DN, greatly exceeding this amplitude. Detection of the Charon-light illuminated terrain could only be done by stacking the complete P\_DEEPIM data set to beat down the shot-noise contribution.}  By averaging all 360 images, the error in the mean level would decrease to 0.47 DN in a single pixel; however, we scaled the simulated noise-level up by $\sqrt{2}$ to 0.66 DN to account for the error in the scattered light model, which was constructed from 360 images with the same background level.  The right panel in Figure \ref{fig:p_mvic} shows the result of adding this level of noise to the binned image on the left.  While nearly all the fine-scale features were lost in the noise, large-scale albedo contrasts, such as between the bright surface of Sputnik Planitia and dark zone of Cthulhu Macula present in the MVIC input image, remain evident in the simulated P\_DEEPIM stack.  The strongest contrast is the sharp delineation of the bright limb of Pluto against the dark background, but unfortunately the strong ring of scattered light from haze in the atmosphere overwhelmed the Charon-light illuminated limb in the real dataset. Figure \ref{fig:cut} presents the simulated trace of the observed flux level moving from the limb to the terminator defined by the Charon illumination.  As we will discuss in $\S\ref{sec:ds_image},$ the simulated trace appears to be in good agreement with the actual intensity trace recovered.

\section{The P\_DEEPIM Charon-Light Images}

\begin{figure}[htbp]
\centering
\includegraphics[keepaspectratio, clip=true, width=6.5in, viewport= 50 50 460 300]{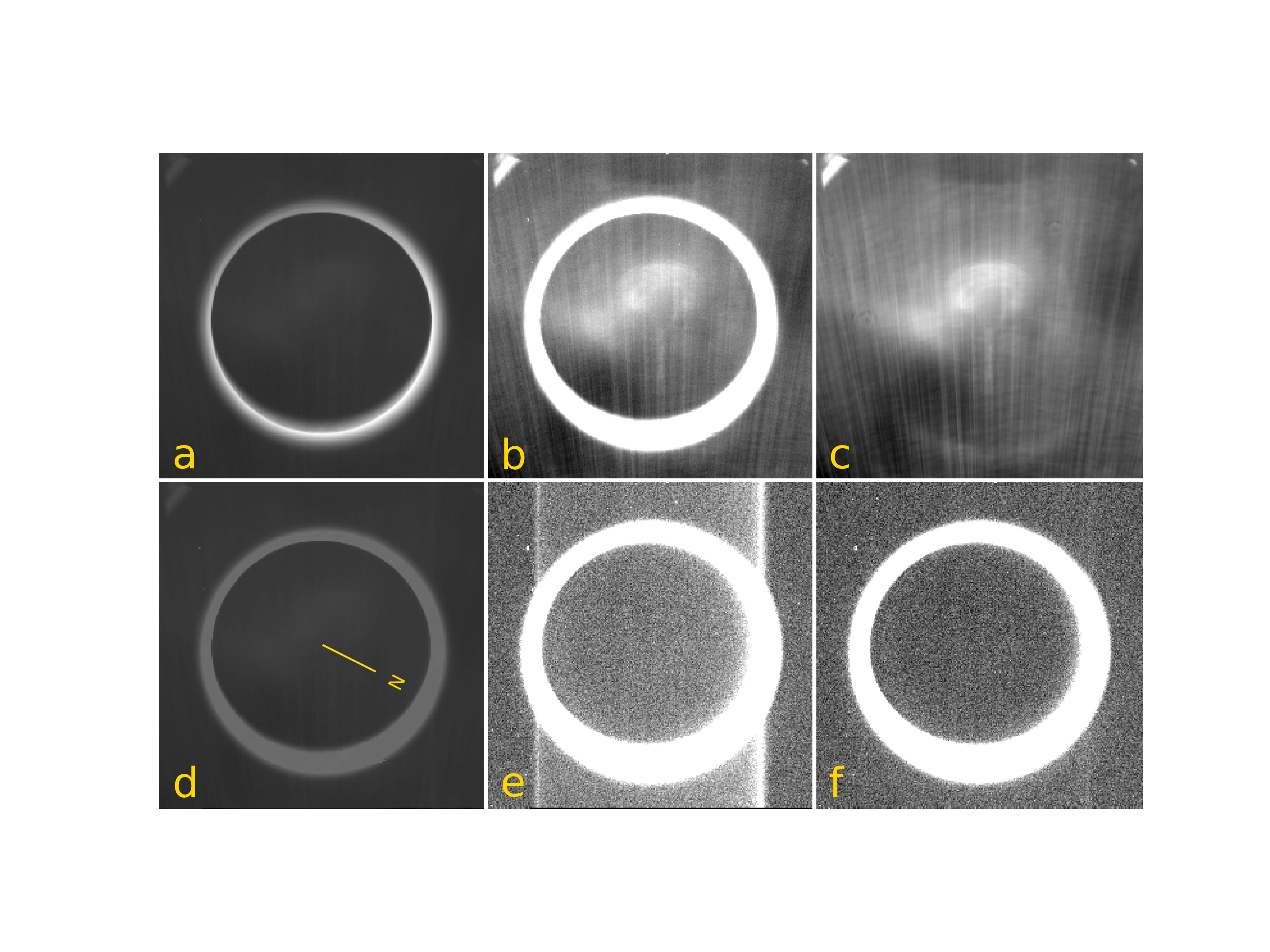}
\caption{The reduction of lor\_0299230814, one of the P\_DEEPIM Pluto images, is shown. a) A logarithmic stretch of the image after repair of its overexposed haze ring.  b) The same image but with a linear range of 600 DN.  Note the highly structured scattered-sunlight background. c) The PCA model of the background. Comparison to b) shows both the high fidelity of the model as well as its greatly reduced random noise. d) The image before the overexposed haze ring has been repaired. The stretch is the same as that in a). e) The {\it un-repaired} image with the background model subtracted. The incomplete charge-smear correction is obvious. The display range is now 100 DN. f) The background subtracted from the repaired image. The display range is the same as in e) The dark side of Pluto in this image (which is bounded by the ring) has now been corrected for both scattered light and charge-smear and can be stacked with the other corrected Pluto images in the  sequence. The spacecraft observed Pluto with the top of the LORRI field rolled to PA $243\adeg3.$ The orientation here is as observed; {the direction to Pluto's north pole is shown in d).}}
\label{fig:example}
\end{figure}

The P\_DEEPIM sequence\footnote{The P\_DEEPIM images are available at the NASA Planetary Data System Small Bodies Node in dataset NH-P-LORRI-3-PLUTO-V3.0 (DOI: 10.26007/6775-8M09).} began at 2015, July 15 01:48 UT (spacecraft time), 14.5 hours after the moment of closest approach, at a range of 693,913 km from Pluto. At that time Pluto still filled most of the LORRI field (description of LORRI is provided by \citealt{lorri} and \citealt{lorri2}). 360 images of Pluto were obtained using LORRI {(MET 0299230808 to 0299231930)} in its $4\times4$-binning mode to maximize sensitivity to faint features (as well as to minimize data volume).  The pixel scale in $4\times4$ mode is $4\asec08$/pixel. The exposure was 0.4s for the first set of 180 images, and 0.3s for the second set. The sequence completed at 02:07 UT, at a of range 709,357 km. The average physical pixel scale at Pluto during the sequence was $\sim14.0$ km. A representative Pluto image is shown in Figure \ref{fig:example}. The spacecraft observed Pluto with the top of the LORRI field rolled to PA $243\adeg3;$ the orientation in the figure reflects that of the observations.

As the Sun was only $14^\circ$ away from Pluto during the sequence, scattered-sunlight strongly contaminated the images. The modeling and subtraction of sunlight in the Pluto images is a major task of the present analysis, which will be discussed in detail in $\S\ref{sec:pca}$.  In anticipation of this problem, however, the P\_DEEPIM program included an additional set of 360 scattered light images offset from Pluto {(MET 0299232248 to 029923341)}, but with the same Sun-spacecraft geometry to provide for modeling and subtraction of the scattered light present in the Pluto images. As with the Pluto images, the first exposure for the first 180 scattered light images was 0.4s, and 0.3s for the second set. During both Pluto segments the pointing was continuously varied in both CCD coordinates to move Pluto around within an $8\amin0\times8\amin0$ box centered in the LORRI field to reduce sensitivity to any particular feature in the camera or the scattered light pattern. The pointing was similarly varied over the background imaging segment.

\subsection{Repair and Reconstruction of the P\_DEEPIM Images}

Despite the short exposure times used for the Pluto images, the bright ring of forward-scattered-sunlight caused by haze particles in the Pluto atmosphere was strongly overexposed. The loss of knowledge of the detailed brightness-distribution of the haze ring prevented proper reduction and interpretation of the images as processed  by the standard LORRI calibration pipeline.  Since LORRI has no shutter mechanism, preparing its raw images for analysis required correcting them for the charge-smearing that occurs along the CCD-columns both before and after the exposure \citep{lorri, lorri2}. In brief, the LORRI CCD  is continually clocked when not integrating during an exposure. When Pluto was positioned on the CCD in advance of an exposure, it contributed charge to all CCD rows as they were clocked through the image of the planet on the chip.  This phenomenon was repeated as the CCD was read out at the end of an exposure.  In general terms, a smeared-out image of Pluto was generated over the full extent of the CCD columns, with the intensity of the smeared-image relative to the desired science-image determined by the length of the science exposure compared to the short interval required to clock the CCD rows through  the Pluto image. 

In a properly exposed {\it raw} image, which includes the nominal science exposure {\it plus} a smeared image of the object, it is possible to recover the science exposure alone \citep{lorri2}. In short, given knowledge of the nominal exposure time relative to the various clocking times of the CCD vertical shift register\footnote{The continual ``scrubbing'' clocking done prior to an exposure and the readout clocking done after it have slightly different frequencies.}, the observed pixel values can be related to the true non-smeared values in a system of linear equations.  This charge smear correction is part of the standard LORRI reduction pipeline.  Unfortunately however, once a pixel is overexposed, knowledge of its true intensity is lost at that location, even though the true input flux contributed fully to the shorter smearing exposures when the CCD was being clocked.  In the case of the P\_DEEPIM sequence, this meant that the standard LORRI pipeline smearing correction was incomplete, leaving smeared light from the bright haze ring behind in the dark nightside disk of Pluto interior to the ring.  This artifact is evident in panel e) of Figure \ref{fig:example}.

The solution was to reconstruct the overexposed portions in each of the P\_DEEPIM images using a 0.15s $1\times1$ mode (unbinned) LORRI image of Pluto (MET 299235359) taken only 32m after completion of the P\_DEEPIM sequence.  The haze ring was correctly exposed in this reference image. The change in the illumination phase from the end of P\_DEEPIM to when the reference image was obtained was only $0\adeg07$ and was ignored. Further, on the assumption that the distribution of haze does not vary rapidly with longitude, the small $1\adeg3$ differential Pluto rotation between the P\_DEEPIM and reference image was also assumed to be unimportant.

This procedure required processing the ``raw" uncalibrated images delivered by the spacecraft with an {\it ad hoc} reduction pipeline, rather than using the standard LORRI pipeline.  This re-reduction also allowed a few other non-standard reduction steps to be performed. The bias-level estimation was done by an algorithm that measures the mode, rather than the median, of the DN values in the CCD dark-column.  A special ``jail-bar" correction was also applied, which accounts for the small bias offset between odd and even CCD columns.  Both corrections are discussed  in \citet{lorri2}.

Use of the reference image required first interpolating it to match each image in the P\_DEEPIM sequence, accounting for Pluto's variable location in the LORRI field and its steadily increasing range over  the sequence.  The next step was to generate a charge-smeared version of the adjusted reference image. Again, the goal was to reproduce the overexposed and charge-smeared portions of the raw P\_DEEPIM images as delivered by the camera. A crucial part of this operation was to first {\it multiply} the reference by the CCD flat field appropriate for the location of the overexposed ring in the given P\_DEEPIM image.  This step was required, as the charge-smearing depends on the true pixel sensitivity at any location in the LORRI field, not the uniform sensitivity established by flat-field calibration.

The smeared reference image was then binned to $4\times4$ mode and patched into the overexposed portions of the target P\_DEEPIM image. At this point the target image in raw format had been repaired. One small adjustment made at this stage was to remove and repair cosmic-ray events, which were not affected by charge-smearing, and thus should not be included in the calculation of the smearing correction.  Next, applying the charge-smearing correction algorithm, and then performing the flat-field correction, generated a calibrated and repaired version of the P\_DEEPIM image.

An example of an image before and after the over-exposure repair is given in panels d) and a) of Figure \ref{fig:example}. Without the repair, the charge-smear correction is clearly incomplete (panel e), but is excellent in the repaired image (panel f). In passing, we note that the importance of the charge-smearing correction for proper reduction of the P\_DEEPIM sequence motivated us to reexamine how this correction is conducted and to develop an improved smearing-correction algorithm \citep{lorri2}. The new algorithm provides both more efficient and significantly better charge-smear correction. 

\section{Correcting for scattered-sunlight}\label{sec:pca}

Correcting for the strong scattered-sunlight in the LORRI images was essential to detecting the Charon-light reflected by Pluto --- the sunlight level is three orders of magnitude stronger than the faint Charon-light.  A strong challenge to its removal was that the pattern of scattered light is highly structured, as can be seen in Figure \ref{fig:example}. The pattern further varied significantly with even small positional shifts of the field. In anticipation of this problem, the P\_DEEPIM sequence included an equal number of background images  that attempted to duplicate the Sun-spacecraft geometry, although again no given Pluto image is likely to find a perfect match in the background set, given the sensitivity of the scattered light pattern to small pointing shifts.

Our solution was to characterize the ensemble of background images with a principal component analysis (PCA) approach, which generated a set of orthogonal ``eigenimages" {(see Figure \ref{fig:eigen_im})} that could be used to represent the domain over which the background varied.  These could then be used to construct a background model for any Pluto image under  the assumption that the scattered-light background in the image was similar to the those in the background set.  Expressed another way, the individual background images were presumed to define the space of background structure, with the basis providing a continuous and complete representation of any point in that space. The basis could thus further represent the background in any Pluto image that fell within or close to this space as a linear combination of the eigenimages.

An important feature of PCA is that the eigenimages can be sorted by the amount of variance over the input image set that they incorporate.  The first eigenimage, roughly speaking, is similar to a simple weighted average for the set.  The second eigenimage accounts for the most important variations about this image, while additional eigenimages provide successively smaller corrections. At some point the increasingly less significant eigenimages begin describing the variance due to noise sources.  In practice, only a small set of eigenimages were needed to account for the significant scattered-light features in the image set.  In the present case, indeed only the first 16 eigenimages were needed, based on a visual evaluation of the quality of the background models.  This represented a compact parametric description of the background in any image and led to a strong suppression of noise in the scattered-light models.

\subsection{Using PCA to Construct Background Models}

In detail, our task was to build a background model image, $\textbf{B}_i,$ for each of $i=1,\ldots,N_P$ images of Pluto, $P_i,$ using PCA on an ensemble of $j=1,\ldots,N_S$ scattered-sunlight images, $\textbf{S}_j.$  In our case, $N_S=N_P,$ but that is not required. This treatment was heavily influenced by \citet{pca} and \citet{pca2}, who gave detailed discussions on the use of PCA for data modeling.  The mathematical formalism presented in this section has been adopted from a more extensive development presented by \citet{pca2}.

\begin{figure}[htbp]
\centering
\includegraphics[keepaspectratio, width=5.0in, viewport= 120 20 350 350] {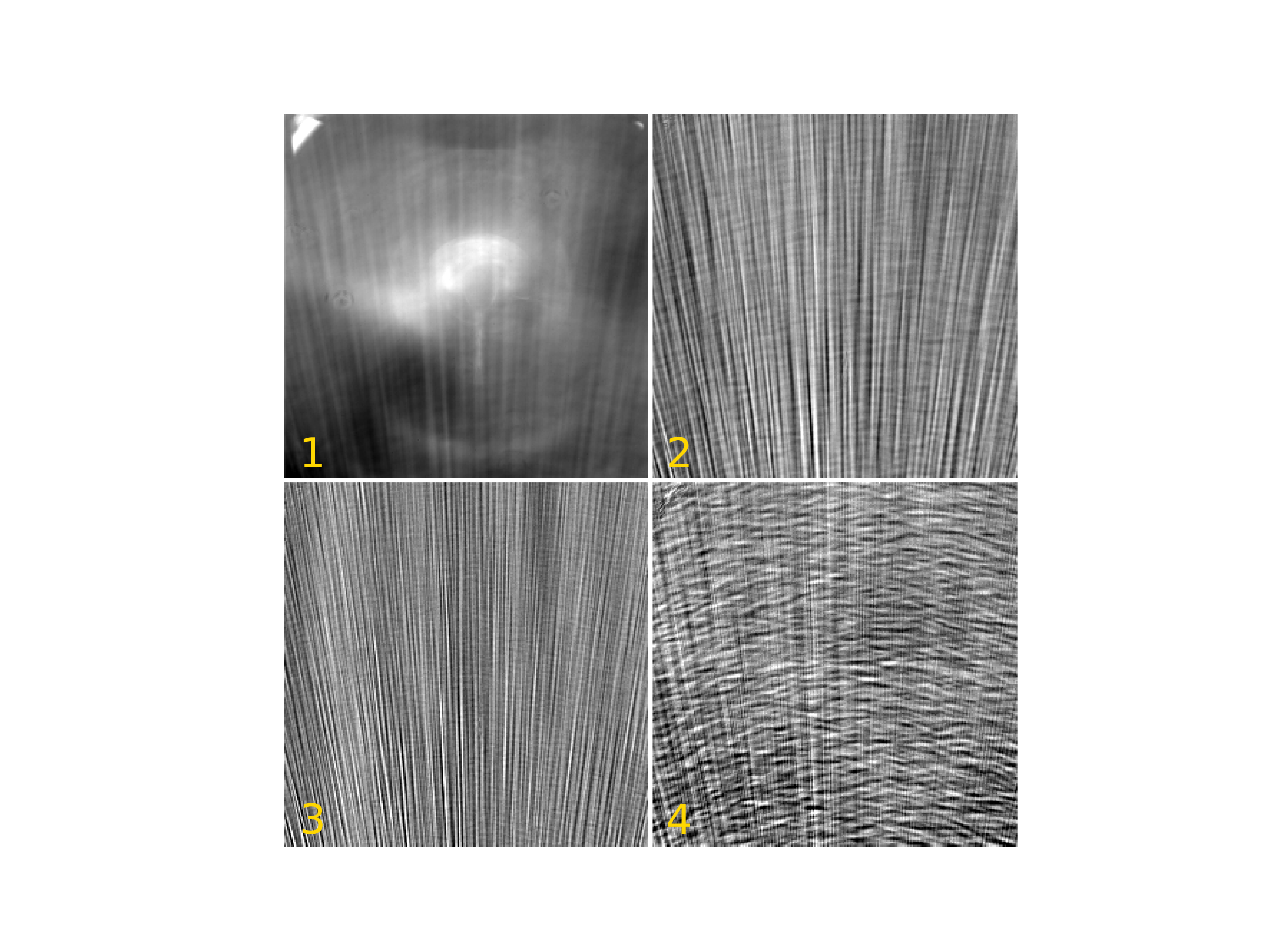}
\caption{The first four scattered-sunlight ``eigenimages" appropriate for lor\_0299230814, are shown. The first eigenimage is approximately the average of all scattered-sunlight images.  The next three images modify its fine structure.  The eigenimages are orthogonal. The background model for lor\_0299230814 is a linear combination of the first 16 eigenimages.  The image stretch is arbitrary for each panel.  The eigenimages are ordered by the amount of variance they account for in the ensemble, and have sequentially decreasing amplitudes in absolute units. }
\label{fig:eigen_im}
\end{figure}

For a given Pluto image, the first step was to define the image domain over which the background would be fitted, which is unique to each Pluto image, given the variable pointing over the sequence. This domain was defined by a mask-image, $\textbf{m}_i,$ which is unity for the background pixels in $\textbf{P}_i$ and zero elsewhere. In detail, the mask is a disk of radius 105 pixels, which is large enough to cover Pluto and its haze ring, once its center was adjusted to track Pluto's location in the field. This ``wandering" mask was applied to all scattered-sunlight images,  yielding the set $\textbf{S}'_{ij}=\textbf{m}_i\textbf{S}_j.$  Note that this procedure meant that we built a basis of scattered-sunlight eigenimages that was unique to each Pluto image.

To construct the basis from the set of $\textbf{S}'_{ij},$ we computed an $N_S\times N_S$ cross-correlation matrix, $C_i$, of all the images in the set, where
\begin{equation}
C_i={A'}_i^{~T} {A'}_i,
\end{equation}
and
\begin{equation}
{A'}_i \equiv [\textbf{S}'_{i1} \,\, \textbf{S}'_{i2} \,\, \cdots \,\, \textbf{S}'_{iN_s}],
\end{equation}
which is an $m\times N_S$ matrix holding each $\textbf{S}'_{ij}$ in a column of length $m,$ the number of pixels in a single image. We then computed the eigenvectors,  $\textbf{v}_{ik},$ and eigenvalues, $\lambda_{ik},$ of $C_i,$ which satisfied the relationship
\begin{equation}
C_i \textbf{v}_{ik} = \lambda_\gamma\textbf{v}_{ik}.
\end{equation}
The eigenvectors have dimension $N_S$.
Each eigenvector has a corresponding eigenimage, 
\begin{equation}\label{eq:eigenimages}
{\textbf{U}'}_{ik}={A'}_i\textbf{v}_{ik};
\end{equation}
again these were the orthogonal scattered-light basis functions appropriate to the specific Pluto image, $\textbf{P}_i.$ Examples from one set of eigenimages are shown in Figure \ref{fig:eigen_im}. The background image for any Pluto image was then generated as a linear combination of {\it unmasked} eigenimages,
\begin{equation}\label{eqn:back}
\textbf{B}_i=\sum_{k=1} ^{N_e} a_{ik}{A}\textbf{v}_{ik},
\end{equation}
where $N_e\leq N_S$ is the number of eigenimages to be used, and the matrix, $A,$ where
\begin{equation}
A \equiv [\textbf{S}_{1} \,\, \textbf{S}_{2} \,\, \cdots \,\, \textbf{S}_{N_s}],
\end{equation}
holds the {\it unmasked} scattered-sunlight images (and is thus the same for all Pluto images). The masked areas, of course, correspond to the disk of Pluto, which is where the background correction is actually needed.  The assumption is that the background for the disk is a valid interpolation from the measured amplitudes of the masked-eigenimages derived from the image area surrounding the disk and haze ring of Pluto.  The coefficients in equation (\ref{eqn:back}) were indeed derived by a dot product of the {\it masked} eigenimages on the image of Pluto,
\begin{equation}
a_{ik} = {\textbf{U}'_{ik}\cdot \textbf{P}_i\over \textbf{U}'_{ik}\cdot \textbf{U}'_{ik}},
\end{equation}
after first applying the small correction of subtracting 1.2 DN from the Pluto images to account for the average level of scattered light from Pluto itself in the unmasked image margins.
\begin{figure}[ht]
\centering
\includegraphics[keepaspectratio, width=7.0in, viewport= 65 50 490 495] {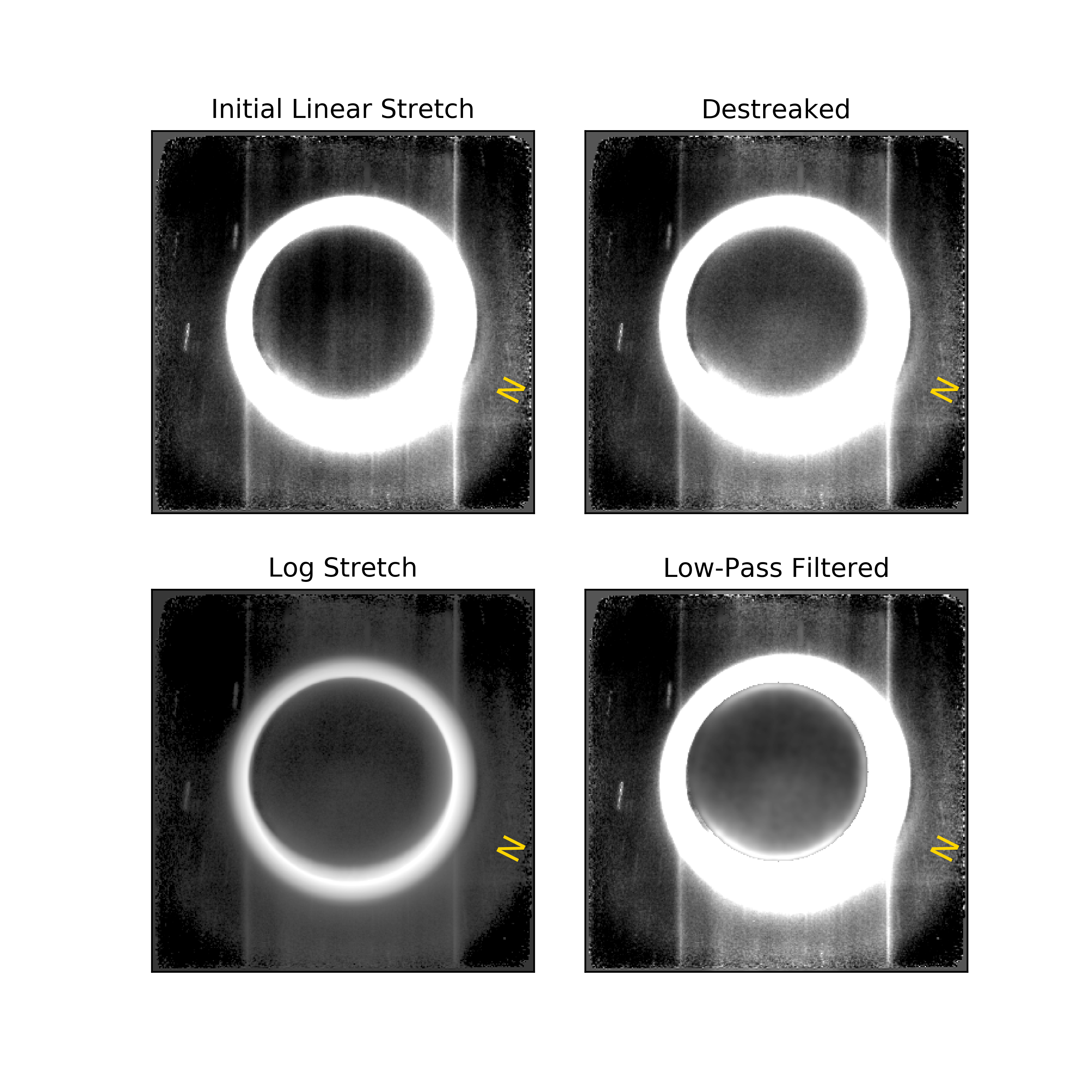}
\caption{The completed stack of the P\_DEEPIM sequence after the scattered-sunlight background has been subtracted. Top left is the initial stack with a display range of 12 DN. Top right is the stack after correcting for vertical streaking. Bottom right is destreaked stack after filtering the disk interior to the haze ring with a low-pass Fourier filter.  Bottom left panel shows a logarithmic stretch of the stack. At its bottom haze extending beyond the illuminated limb remains in sunlight, and creates a crescent of apparent illumination. The spacecraft observed Pluto with the LORRI field rolled by $243\adeg3.$ {The direction to Pluto's north pole is indicated by `N.'}}
\label{fig:dark_image}
\end{figure}
An example background image is shown in c) in Figure \ref{fig:example}. The PCA-generated image clearly matches all the fine details in the corresponding Pluto image, yet it has greatly reduced random noise.  Subtraction of the background image effects an excellent elimination of the scattered-sunlight pattern in the Pluto image.

\clearpage

\section{An Image of the Dark Side of Pluto}\label{sec:ds_image}

Once the background images were generated and subtracted from each of the Pluto images, it was simple to interpolate the residual images to a common range (that of the first image in the sequence) and compute an average residual image over the sequence.  The resulting stack is shown in Figure \ref{fig:dark_image}.  The upper-left panel shows the initial average of the residual images.  A small amount of residual smeared charge remained, but the image also had fine-scale vertical streaks.  The upper-right panel shows this image after correcting for the streaking as follows. A median intensity was computed for each image column within the dark disk of Pluto.  The vector of medians was fitted with a spline to preserve large-scale intensity trends and isolate only the high-frequency components of the streaking. The bottom-left panel shows a logarithmic stretch of this image to visualize the smooth transition of the bright haze ring into the disk of Pluto.  The final lower-right panel shows the disk within the upper-right panel processed with a low-pass filter to remove noise and better isolate large-scale structure in the Charon-light illuminated terrain. 

\begin{figure}[thbp]
\centering
\includegraphics[keepaspectratio, width=7.0in, viewport= 60 50 520 495] {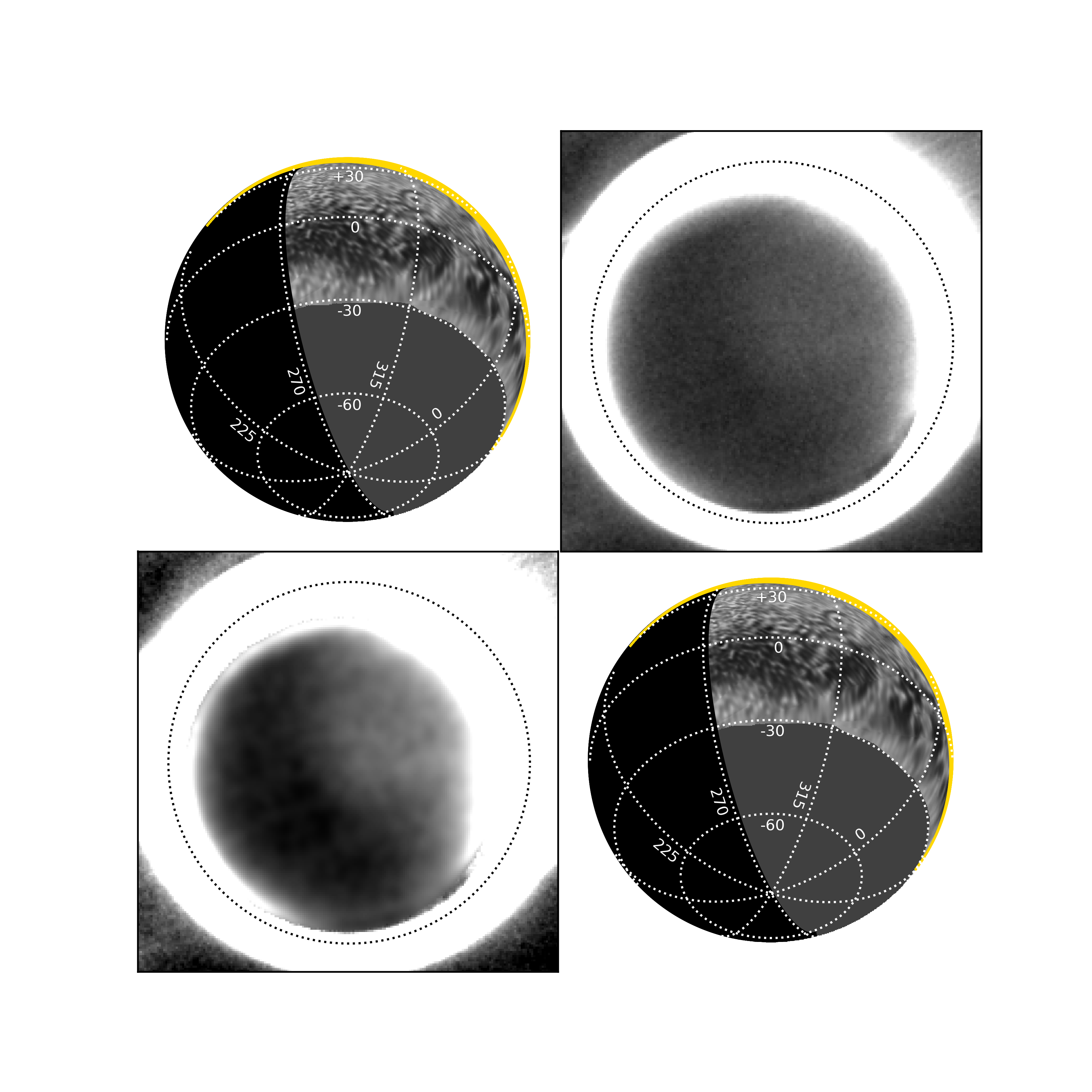}
\caption{The stack presented in Figure \ref{fig:dark_image} now rotated to place north at the top of the figure.  The dashed circle shows the location of Pluto's limb. The graphical representation presented in Figure \ref{fig:pluto_geo} is repeated in the upper left and lower right, scaled to match the size of Pluto's disk in the stack. The destreaked stack in the upper-right panel of Figure \ref{fig:dark_image} is repeated in the upper-right panel of this figure. The low-pass filtered stack in Figure \ref{fig:dark_image} is repeated in the lower-left of this figure, but with a harder stretch to better highlight the faint pattern of Charon-light illuminated terrain.  The candidate bright basin is at the center of the disk. The low-albedo south polar region is evident as the dark area right of the Charon-light terminator and at the bottom of the disk.}
\label{fig:map_comp}
\end{figure}

 Figure \ref{fig:map_comp} shows the final Charon-light image (with north now at the top) compared to the pre-existing terrain map, repeated from Figure \ref{fig:pluto_geo}. The Charon-light image is darkest over a zone that corresponds to longitudes west of $270^\circ$ that receives neither sunlight nor Charon-light. The boundary between it and the Charon-illuminated eastern portion of the images corresponds nicely to the position of the Charon-light terminator. This concordance gives us confidence that we have detected Charon-light illumination. The strength of the Charon-light illumination also appears to agree with our quantitative expectations of its strength.  Figure \ref{fig:cut} shows an intensity trace measured (red-dotted) from a 20-pixel-wide cut along the vertical diameter of the ``destreaked" stack shown in Figure \ref{fig:dark_image}. The observed trace does show the strong effects of the haze ring (and the narrow zone of full sunlight), unlike the case for the simulated image; however, its intensity level over the center of the disk and its decrease into the zone unilluminated by Charon-light is in good agreement with the simulated intensity trace. The terrain sampled by the P\_DEEPIM and simulated traces will be different, of course; thus differences between the two are expected. The overall amplitude and gradients in the two traces are set by the illumination pattern, however, and these are in agreement. We do note one caveat, which is that the scattered light was slightly oversubtracted by $\sim1.5$ DN.  In Figure \ref{fig:cut} this bias has been removed so that the observed trace does go to zero in the unilluminated zone.
 
In passing, we note that the stacked Charon-light image provides the deepest exposure of the atmospheric haze ring obtained during the encounter. The haze is visible out to $262\pm14$ km from the limb of the planet. The ``log stretch" panel in Figure \ref{fig:dark_image} also shows a zone beyond the sunlight terminator (at the bottom of the image) where the haze remains in direct sunlight, while the underlying terrain is in twilight. This zone extends $33^\circ$ away from the geometric limb, consistent with the radial extent of the haze (see \citealt{cheng} for a detailed discussion of the structure of the atmospheric haze).
 
\begin{figure}[thbp]
\centering
\includegraphics[keepaspectratio, width=5.0in]{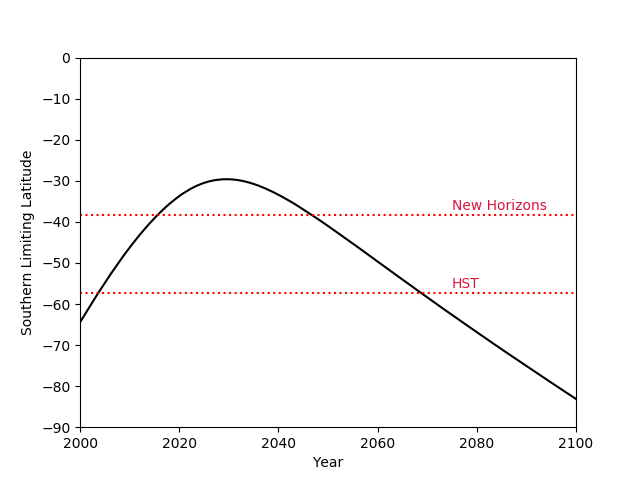}
\caption{The southern limit of latitudes on Pluto receiving direct sunlight are shown as a function of time over the present century.  New Horizons encountered Pluto in mid-2015, while the \citet{buie} map of Pluto was generated from images obtained in 2002-3.  At present, the Sun is still moving north, and it will still be three decades before it illuminates terrain further south than the limit at the time of the encounter. The ability to map terrain illuminated by Charon-light or Plutonian twilight will still be important for missions returning to Pluto for most of the century remaining \citep{howett}.}
\label{fig:limit}
\end{figure}

\begin{figure}[htbp]
\centering
\includegraphics[keepaspectratio, width=7.0in,viewport= 60 30 520 280]{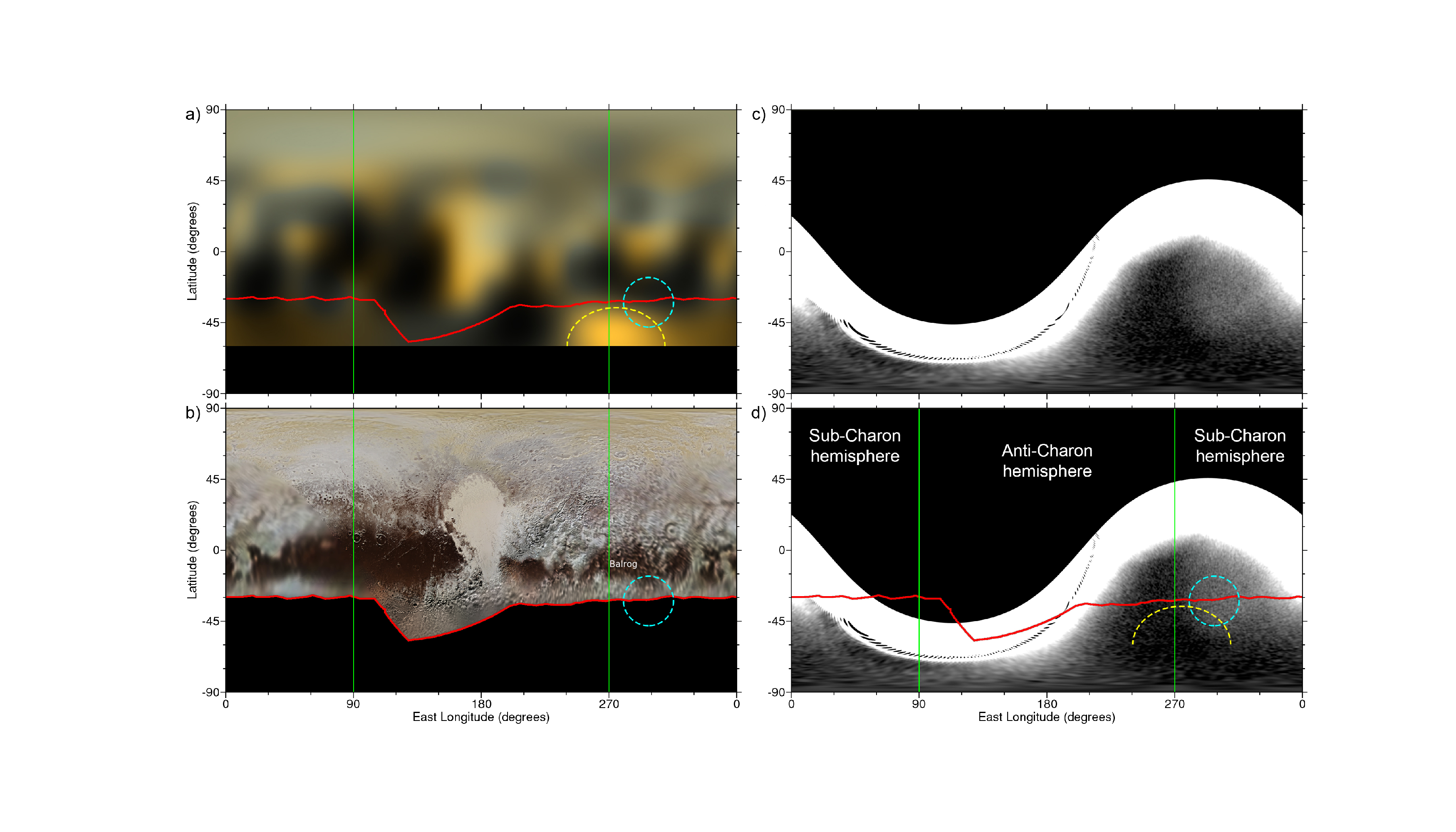}
\caption{Various global maps of Pluto shown in simple cylindrical projection and centered at the anti-Charon longitude of $180^\circ$ E. (a) The synthesized HST map of Pluto from \citet{buie}. (b) Colorized New Horizons MVIC and LORRI mosaic. (c) The destreaked Charon-illuminated stack from Figure \ref{fig:map_comp}, shown to approximately the same stretch, and (d), the same as (c), but annotated.  On the maps in (a), (b), and (d), the red line indicates the southern boundary of the New Horizons imaging in the mosaic in (b), the vertical green lines indicate the location of the Charon-light terminator that separates the anti-Charon hemisphere from the sub-Charon hemisphere, and the dashed blue ellipse indicates the approximate boundary of the bright region identified in the Charon-light imaging. {The bright region is SE of Balrog Macula (labeled in panel b).} The dashed yellow semi-ellipse in (a) and (d) indicates the approximate boundary of the bright region identified in the HST imaging.}
\label{fig:cyl}
\end{figure}

 \subsection{The Bright Southern Mid-Latitude Region}
 
 Figure \ref{fig:cyl}c presents the destreaked stack from Figure \ref{fig:map_comp} in simple cylindrical projection, with an annotated version shown in Figure \ref{fig:cyl}d. Some variability in brightness is apparent in the Charon-illuminated portion of the map (i.e., the sub-Charon hemisphere). A relatively bright region is centered at $\sim33^\circ$ S, $298^\circ$ E and extends from $15^\circ$ to $50^\circ$ S and from $\sim280^\circ$ to $315^\circ$ E, measuring $\sim635$ km in diameter (its approximate boundary is indicated by a dashed blue outline). Terrain toward the equator appears slightly darker (and corresponds well to the dark equatorial band seen in approach imaging in Figure \ref{fig:cyl}b). The terrain over the south polar region appears darker than either the mid-latitude or equatorial regions and will be discussed in the next subsection.
 
 Observations made by the Hubble Space Telescope (HST) in 2002-3 \citep{buie} revealed some of Pluto’s southern terrain (within $30^\circ$ -- $60^\circ$ S) that was in winter darkness during the 2015 reconnaissance by New Horizons (Figure \ref{fig:limit}). Figure \ref{fig:cyl}a reproduces the synthesized HST map of Pluto from \citet{buie}, and Figure \ref{fig:cyl}b shows the New Horizons color map of Pluto. Through comparison of the two maps, it can be seen that the bright Sputnik Planitia, the dark latitudinal zone just south of the equator, and the intermediate-albedo polar latitudes are all evident in the HST imaging, which additionally reveals a bright province in the southern hemisphere centered at $\sim30^\circ$ S, $275^\circ$ E, and that has a similar albedo and physical extent to Sputnik Planitia. It extends from $\sim240^\circ$ to $310^\circ$ E. 
 
 As seen in Figure \ref{fig:cyl}a, the bright region in the Charon-light imaging is located at the northeastern boundary of the bright region in the HST imaging, and as seen in Figure \ref{fig:cyl}b, its boundaries also extend into the farside approach imaging of the New Horizons color map.  While they only partially overlap, the proximity of the bright regions seen in the Charon-light and HST imaging may suggest that they are representative of a single bright region located in the southern mid-latitudes of Pluto’s farside.  The bright region in the Charon-light imaging may extend as far west as the bright region in the HST imaging, but is truncated by the Charon-light terminator at $270^\circ$ E.  Close inspection of the farside approach imaging in Figure \ref{fig:cyl}b does show a contradiction between it and the HST map near the latter's resolution limit.  The present bright region appears to correspond to an area of increased albedo in the New Horizons map south of the dark terrain designated as Balrog Macula, while the same area in the HST map appears to be an extension of Balrog Macula.
 
 If the southern bright region seen in New Horizons and HST images is a genuine surface feature, then it may represent a regional-scale concentration of ${\rm N_2}$-rich or ${\rm CH_4}$-rich surface volatile-ices. Given that the present bright patch is centered at a higher latitude than Sputnik Planitia ($\sim33^\circ$ S, as opposed to $25^\circ$ N), and so experiences higher average insolation \citep{ham16, earle}, it is likely that such a concentrated deposit would require a deep basin to ensure its thermodynamic stability, as is the case for Sputnik Planitia \citep{bert16, bert18}. Alternatively, the concentration of volatiles could correspond to a seasonal or perennial deposit similar to the mid-latitudinal band observed by New Horizons in the northern hemisphere \citep{schmitt, proto}, and modeled in \citet{bert19}. In this regard, the bright region could represent a seasonal extension/evolution of the feature seen in the HST map since the 2002-3 epoch of the \citet{buie} observations. High topographic relief at these southern mid-latitudes in the eastern hemisphere could prevent the deposit from forming a continuous band extending around Pluto as in the northern hemisphere. Lastly as an alternative, high-altitude mountains at these southern latitudes could favor ${\rm CH_4}$ condensation on their peaks (as seen in Cthulhu, \citealt{moore, bert20b}) creating a brighter surface than the surroundings.  

 \subsection{A Dark South Polar Region}
 
 Based on the geology observed in Pluto’s encounter and anti-encounter hemispheres \citep{moore, farside}, and expectations from climate modeling \citep{bert19, bert20, johnson}, it would be reasonable to expect that the mid- to high-latitudes in the southern hemisphere would generally display fairly old, intermediate-albedo mantled terrains, as in the northern hemisphere, with perhaps some scattered patches of ponded nitrogen-ice at the lower latitudes.  However, the different seasonal durations experienced by the northern and southern hemispheres arising from Pluto’s highly elliptical solar orbit could produce seasonal asymmetries between northern and southern ice formations \citep{earle}.  Again, using the {numerical} climate models cited above, we may speculate at the likely surface volatile ice coverage that exists in Pluto’s dark southerly latitudes (i.e., that spread across uplands terrain and disregarding any concentrations within basins). Based on the following lines of reasoning, the climate modeling does imply a southern hemisphere that is on average darker than the northern hemisphere, which is broadly consistent with our results. 
 
 First, ${\rm N_2}$ ice is not expected to cover as large an area in the southern hemisphere as it covers in the northern hemisphere. Otherwise, the surface pressure would have significantly dropped prior to 2015, which is inconsistent with New Horizons observations and recent stellar occultation results \citep{stern, gladstone, sicardy, meza, johnson, young21}.
 
 Second, summer in the southern hemisphere ended about three decades prior to the New Horizons encounter. It can be expected that significant fractions of ${\rm N_2}$ and ${\rm CH_4}$ ice had recently sublimated in the south, which would lead to a relatively dark southern surface in 2015 as a darker substrate was revealed, or a dark refractory lag deposit accumulated. Although volatile ice would be condensing in the southern hemisphere in 2015 (although the condensation rates would be relatively slow, as in 2015 the season was fall, not winter), the net accumulation would not yet be enough to replace what had been lost during the previous summer, according to the {numerical climate} models {\citep{bert19}}. The amount of volatile ice in the southern hemisphere in 2015 would have been still close to its annual minimum, {or even absent at some locations}, which could explain the relatively dark surface of the southern hemisphere. 
 
 Lastly, the albedo of the southern hemisphere could also have changed over time owing to deposition of organic haze material \citep{grundy}. It can be expected that the darkening of the surface in the southern hemisphere due to the haze deposit was at a maximum at the end of southern summer, as both volatile ice sublimation and haze production and deposition are maximal above the southern hemisphere and pole during this season. This haze darkening could also increase the surface temperature and prevent further volatile ice deposition.
 
 {Against the present results, however, we note that \citet{young1} and \citet{young2, young3} constructed albedo maps of the Pluto over its Charon-facing hemisphere, using Pluto-Charon mutual transiting events over 1985-1990 as their input data. Their map reconstructions consistently recover {\it bright} south polar regions; also see the extensive discussion on the brightness of the south polar region in \citet{young21}.  As the epoch of observations corresponds to the end of the southern summer, this would appear to contradict our suggestion that the south pole darkened over the duration of its summer season.}
 
 \section{Reflections on the Dark Side}
 
 We have successfully recovered an image of Pluto's dark hemisphere  under faint partial illumination by sunlight reflected off of its moon Charon. The technical reduction of the imagery required solutions to two major problems.  The first task was effecting removal of the substantial smeared-charge  signal present in the LORRI images, due to information lost in the overexposed haze rings in the Charon-light images. The solution required both redeveloping the smeared-charge correction algorithm, as well as reconstructing the overexposed portions of the images with higher-resolution LORRI images taken close in time to the P\_DEEPIM sequence. The second task was constructing high-fidelity model images of the strong and highly-structured scattered-sunlight background present in the images due to Pluto's small elongation from the Sun during departure. This task benefited from a large set of scattered light images taken at the same Sun-spacecraft geometry as was used for the Charon-light images. Principal components analysis was used to construct background models unique to each Charon-light image.
 
 The final stacked Charon-light image is noisy owing to the strong shot noise from the scattered-sunlight background, and thus cannot provide useful information on fine angular scales. Still, the image appears to reveal a southern bright region, which we argue may be due to ${\rm N_2}$ or ${\rm CH_4}$ ice deposits.  The south polar region also appears to have significantly lower albedo than the rest of the Charon-light illuminated zone.  This is qualitatively consistent with the HST map of \citet{buie}, which shows low albedos at its southern latitude limit of $\sim-60^\circ,$ but the present map extends further down to the pole itself. A presently low-albedo south pole might be expected, for example, as a consequence of the recently completed southern summer, during which ${\rm N_2}$ surface ice would have sublimed, and dark haze particles would have been preferentially produced and settled out over the polar region.  While ${\rm N_2}$ would be condensing in the south now, the buildup has not continued long enough to significantly alter the albedo of the pole.
 
 We do note that our interpretation of the Charon-light image is based on a simple visual evaluation of the large-scale features recovered.  A more sophisticated approach might be had by forward-modeling various input albedo patterns, fully accounting for the nonuniform illumination provided by Charon over the sub-Charon hemisphere, and the material reflectivities as a function of illumination phase of Pluto's surface.  We are currently evaluating whether such an approach might allow tighter constraints on the properties of the Charon-light illuminated terrain observed by the P\_DEEPIM sequence.
 
 Verifying the Charon-light image, as well as testing our hypotheses for the origin of the features seen, could be done by a follow-on mission to Pluto, adaptive optics on one of the 25m-class ground-based telescopes now under development, as well as one of the next-generation space telescopes now under study, such as LUVIOR \citep{luvior}. All Earth-based instruments, however, can only image the dayside hemisphere of Pluto. Figure \ref{fig:limit} shows that for the next several decades Charon-light and Plutonian-twilight imaging will still be required for imaging the southern latitudes not in sunlight at the time of the New Horizons encounter.  These considerations, for example, are reflected in the design of the Persephone Pluto orbiter concept, which includes a low-light-level camera, and reaction wheel pointing control, allowing for long integrations \citep{howett}.  An orbiting spacecraft will also have excellent control of the imaging geometry, being able to avoid the scattered-sunlight and bright haze ring that affected the present imaging.  What was tricky for New Horizons should be easy for Persephone.
 
\acknowledgments

We thank NASA for their funding and continued support of the New Horizons mission. The data presented here were obtained during the New Horizons exploration of the Pluto/Charon system.

{}

\end{document}